\documentclass[showpacs,prb,twocolumn,aps,amsmath,amsfonts]{revtex4}
\usepackage{psfrag} \usepackage{epsfig}

\newcommand{\tsc}[1]{\,\text{tsc} \,#1\,\,}

\newcommand{\bs}{\textbf}

\begin{document}
\title{Control of Dephasing and Phonon Emission in Coupled Quantum
  Dots} 
\author{S. Debald $^1$, T. Brandes $^2$, and B. Kramer $^1$}
\affiliation{ $^1$ University of Hamburg, 1. Inst. Theor. Physik,
  Jungiusstr. 9, 20355 Hamburg, Germany} 
\affiliation{ $^2$ Department
  of Physics, University of Manchester Institute of Science and
  Technology (UMIST), P.O. Box 88, Manchester M60 1QD, United Kingdom}
\date{\today{ }}
\begin{abstract} 
  We predict that phonon subband quantization can be detected in the
  non-linear electron current through double quantum dot qubits embedded into
  nano-size semiconductor slabs, acting as phonon cavities. For particular
  values of the dot level splitting $\Delta$, piezo-electric or deformation
  potential scattering is either drastically reduced as compared to the bulk
  case, or strongly enhanced due to phonon van Hove singularities. By tuning
  $\Delta$ via gate voltages, one can either control dephasing, or strongly
  increase emission into phonon modes with characteristic angular
  distributions.
\end{abstract}

\maketitle

Coupled semiconductor quantum dots are candidates for controlling quantum
superposition and entanglement of electron states. The feasibility of such
`qubits' depends on the control of dephasing due to the coupling to low-energy
bosonic excitations of the environment. For example, the electronic transport
thorouhg double quantum dots is determined by the spontaneous emission of
phonons even at very low temperatures \cite{FujetalTaretal}. If two dots are
coupled to each other and to external leads, Coulomb blockade guarantees that
only one electron at a time can tunnel between the dots and the leads.
Dephasing in such a `pseudo spin'-boson system \cite{Legetal87,BK99} is
dominated by the properties of the phonon environment.

As a logical step towards the control of dephasing, the control of vibrational
properties of quantum dot qubits has been suggested \cite{FujetalTaretal}.
Recently, considerable progress has been made in the fabrication of
nano-structures that are only partly suspended or even free-standing
\cite{CR98,BRWB98}. They considerably differ in their mechanical properties
from bulk material. For example, phonon modes are split into subbands, and
quantization effects become important for the thermal conductivity
\cite{SW92,GRKV97,RK98}. The observation of coherent phonons in dots
\cite{KW97} in nanotubes \cite{DE00} are other examples of low dimensional
mesoscopic systems where phonons become experimentally controllable and are
the objects of interest themselves.

Double quantum dots are not only tunable phonon emitters \cite{FujetalTaretal}
but also sensitive high-frequency noise detectors \cite{AK00}. Together with
their successful fabrication within partly free-standing nanostructures
\cite{Blietal00}, this suggests that they can be used to control both
electrons and phonons on a microscopic scale. This opens a path for realizing
mechanical counterparts of several quantum optical phenomena, as for instance
the generation of non-classical squeezed phonon states \cite{HN97} by
time-dependent or non-linear interactions with the electrons.

In this paper, we demonstrate that phonon confinement can be used to gain
control of dissipation in double quantum dots, leading to a considerable
reduction of phonon-induced decoherence.  More precisely, we show that
inelastic scattering and the inelastic current channel for electron transport
in the Coulomb blockade regime can be drastically reduced as compared to a
bulk environment when double dots are hosted by a semiconductor slab that acts
as a phonon cavity.  This suppression occurs at specific phonon energies
$\hbar\omega_0$ when the level splitting is tuned to $\Delta=\hbar\omega_0$.
Furthermore, for larger energy differences $\varepsilon$ between the two dot
ground states, typical properties \cite{MM64Aul73} of a nano-size slab such as
phonon-subband quantization can be detected in the staircase-like electronic
current $I(\varepsilon)$ though the dots. In addition, and very strikingly,
for certain wave vectors we find negative phonon group velocities and phonon
van Hove singularities close to which one can strongly excite characteristic
phonon modes with specific emission patterns.

\begin{figure}[ht]
\epsfig{file=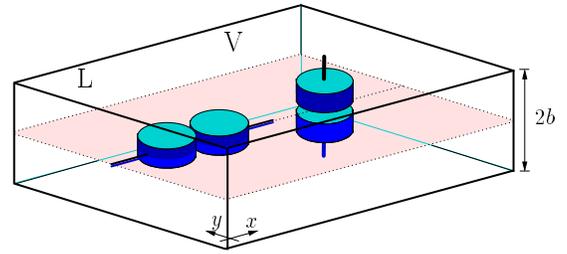,width=0.47\textwidth}
\caption[]{\label{polar}
  Scheme for {\em lateral} (L) and {\em vertical} (V) configurations of a
  double quantum dot in a phonon nano-cavity.}
\end{figure}  

The controlled enhancement or reduction of spontaneous light emission from
atoms is well-known in cavity QED \cite{cavityQED}. Here a single, confined
photon mode can be tuned on or off resonance with an atomic transition
frequency. In contrast to cavity QED, the vanishing of spontaneous emission in
phonon cavities is due to real zeros in the {\em phonon deformation potential
  or polarization fields} rather than gaps in the density of states. This is a
peculiar consequence of the boundary conditions for vibration modes which lead
to complicated nonlinear dispersions even for homogeneous slabs. As a result,
phonon cavities confined in only one spatial direction support a continuous
spectrum, and yet suppression of spontaneous emission is possible.

As a model, we consider two tunnel-coupled quantum dots embedded in an
infinite semiconductor slab of thickness $2b$ in a vertical or a
lateral configuration, Fig. \ref{polar}. The dots are weakly coupled
to external leads, and we assume that both the energy difference
$\varepsilon$ and the coupling strength $T_c$ between the dots can be
externally tuned by gate voltages. In the Coulomb blockade regime, we
adopt the usual description in terms of three many body ground states
\cite{BK99,SN96} $|0\rangle$, $|L\rangle$, and $|R\rangle$ that have
no or one additional electron in either of the dots, respectively. The
coupling to the phonon environment of the slab is described by an
effective spin-boson Hamiltonian
\begin{eqnarray}
  \label{eq:spinboson}
H&=&\frac{\varepsilon}{2}\sigma_z + T_c\sigma_x +
\sum_{\bf{q}}\hbar\omega_{\bf{q}}a_{\bf{q}}^{\dagger}a_{\bf{q}}\nonumber\\ 
&+& \sum_{\bf{q}}\left(\alpha_{\bf{q}} n_L + \beta_{\bf{q}} n_R
\right) \left(a_{\bf{q}} + a_{-\bf{q}}^{\dagger} \right), 
\end{eqnarray}
with $\sigma_z = |L\rangle \langle L| - |R\rangle \langle R|$,
$\sigma_x = |L\rangle \langle R| + |R\rangle \langle L|$, $n_i =
|i\rangle \langle i|$ and $\alpha_{\bf{q}}$($\beta_{\bf{q}}$) the
coupling matrix element between electrons in dot $L$($R$) and phonons
with dispersion \nolinebreak$\omega_{\bf{q}}$.

The stationary current can be calculated by using a master equation
\cite{BK99} and considering $T_c$ as a perturbation. We
consider weak electron--phonon coupling and calculate the inelastic
scattering rate
\begin{eqnarray}
  \label{eq:eprate}
\gamma(\omega)=2\pi  \sum_{\bf{q}}T_c^2 \frac{|\alpha_{\bf{q}} -
  \beta_{\bf{q}}|^2}{\hbar^2 \omega^2}\delta(\omega-\omega_{\bf{q}}).
\end{eqnarray}
For $\hbar\omega= (\varepsilon^2+4T_c^2)^{1/2}$ this is  the rate for
spontaneous emission at zero temperature due to electron transitions
from the upper to the lower hybridized dot level.

On the other hand, in lowest order in $T_c$, the total current
$I(\varepsilon)$ can be decomposed into an elastic Breit-Wigner type
resonance and an inelastic component $I_{\rm in}(\varepsilon) \approx
-e\gamma(\varepsilon)$, where $-e$ is the electron charge.  The double
dot, supporting an inelastic current $I_{\rm in}(\varepsilon)$,
therefore can be regarded as an analyzer of the phonon system
\cite{FujetalTaretal}. One can also consider the double dot as an
emitter of phonons of energy $\hbar \omega$ at a tunable rate
$\gamma(\omega)$.  We show below how the phonon confinement within the
slab leads to steps in $I_{\rm in}(\varepsilon)$ and tunable strong
enhancement or nearly complete suppression of the electron-phonon
coupling.

\begin{figure}[tbh] 
\begin{center}
\epsfig{file=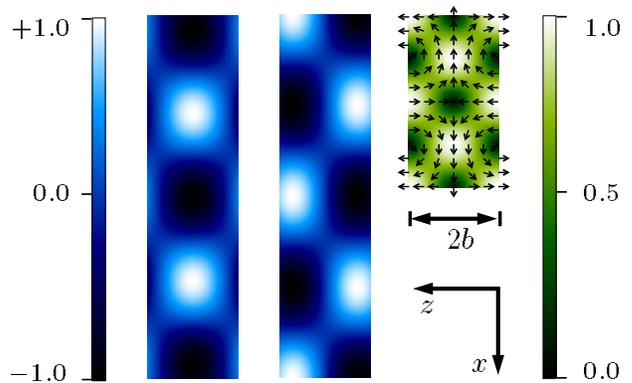,width=0.47\textwidth}
\end{center}
\caption{Deformation potential
  induced by dilatational (left) and flexural modes (center) at
  $q_{\|}b=\pi/2$ ($n=2$ subbands).  Right: displacement field
  $\bs{u}(x,z)$ of $n=0$ dilatational mode at $\Delta=\hbar \omega_0$.
  Greyscale: moduli of deformation potentials (left) and displacement
  fields (right) (arb.  units).}
\label{fields}
\end{figure}

We describe phonons by a displacement field $\bf{u}(\bf{r})$ which is
determined by the vibrational modes of the slab \cite{LL65}. For the
following, it is sufficient to consider dilatational and flexural modes (Lamb
waves). The symmetries of their displacement fields differ with respect to the
slab's mid-plane. They either yield a symmetric elongation and compression
(dilatational mode, Fig.~2\,left) or a periodic bending associated with an
antisymmetric field (flexural mode, Fig.~2\,center). The confinement leads to
phonon quantization into subbands.  For each in-plane component $\bf{q}_{\|}$
of the wave vector there are infinitely many subbands, denoted by $n$, related
to a discrete set of transversal wavevectors in the direction of the
confinement. Since there are two velocities of sound in the elastic medium
associated with longitudinal and transversal wave propagation, $c_l$ and
$c_t$, there are also two transversal wavevectors $q_l$ and $q_t$.  This is in
contrast to the isotropic bulk where one can separate the polarizations. For a
slab, the boundary conditions at the surface lead to coupling between
longitudinal and transversal propagation \cite{MM64Aul73}.

We have numerically determined the solutions $q_{l,n}(q_{\|})$ and
$q_{t,n}(q_{\|})$ of the Rayleigh-Lamb equations that describe the
dynamics of the confined phonons,
\begin{eqnarray} \label{rl1}
\frac{\tan q_{t,n} b}{\tan q_{l,n} b} &=& -\left[ \frac{4 q_{\|}^2
    q_{l,n} q_{t,n}}{(q_{\|}^2  - q_{t,n}^2)^2} \right]^{\pm 1}
\nonumber\\ \omega_{n,q_{\|}}^2 &=& c_l^2 (q_{\|}^2 + q_{l,n}^2) =
c_t^2 (q_{\|}^2  + q_{t,n}^2), 
\end{eqnarray}
together with the dispersion relations $\omega_{n,q_{\|}}$ and the
displacement field \cite{BAMS9594} associated with a confined phonon in
mode ($n, \bf{q}_{\|}$). The exponents $\pm 1$ correspond to
dilatational and flexural modes, respectively.

The contribution $\gamma_n$ of subband $n$ to the
rate (\ref{eq:eprate}) is
\begin{equation}
  \label{eq:gamman}
   \gamma_n(\omega)=\sum_{\bf{q}_{\|}}\!
\frac{|\lambda_{\rm dp/pz}^{\pm}(\bs{q}_{\|}, n)|^2}{\hbar^2 \omega^2}
\! \left|  \alpha \pm e^{i \bf{q}_{\|}\bf{d}} \right|^2  \! \delta(\omega -
\omega_{n,q_{\|}})
\end{equation}
where the vector ${\bf d}$ connects the dots and $\lambda$ is the
coupling strength of the electron-phonon interaction. We assume that
the electron density is sharply peaked near the dot centers and that
the dots are located symmetrically within the slab.  We consider both
the deformation potential (DP), $\alpha= -1$ in (\ref{eq:gamman}), and
piezo-electric (PZ) interaction, $\alpha= +1$. The
coupling strength for DP is
\begin{equation}
\label{lambdadp}
\lambda_{\rm dp}^{\pm}(q_{\|},n) = B_n^{\rm dp}(q_{\|}) ( q_{t,n}^2 -
q_{\|}^2)(q_{l,n}^2 + q_{\|}^2) \tsc{q_{t,n} b},
\end{equation}
where $B_n^{\rm dp} = F_n(\hbar\Xi^2/2\rho\,
\omega_{n,q_{\|}}A)^{1/2}$, $\tsc{x} = \sin x$ or $\cos x$ for
dilatational and flexural modes, respectively, $\Xi$ is the DP
constant, $\rho$ the mass density, $A$ the area of the slab, and $F_n$
normalizes the $n^{\rm th}$ eigenmode.

First, we discuss the deformation potential interaction in the
vertical configuration (Fig.~\ref{polar}), ${\bf{q}_{\|}\bf{d}}=0$,
where only flexural modes couple to the electrons, whereas
dilatational modes lead to a symmetrical deformation field
(Fig.~\ref{fields}\,left) and yield the same energy shift in both of
the dots which does not affect the electron tunneling.

\begin{figure}[tbh] 
  \begin{center}
\epsfig{file=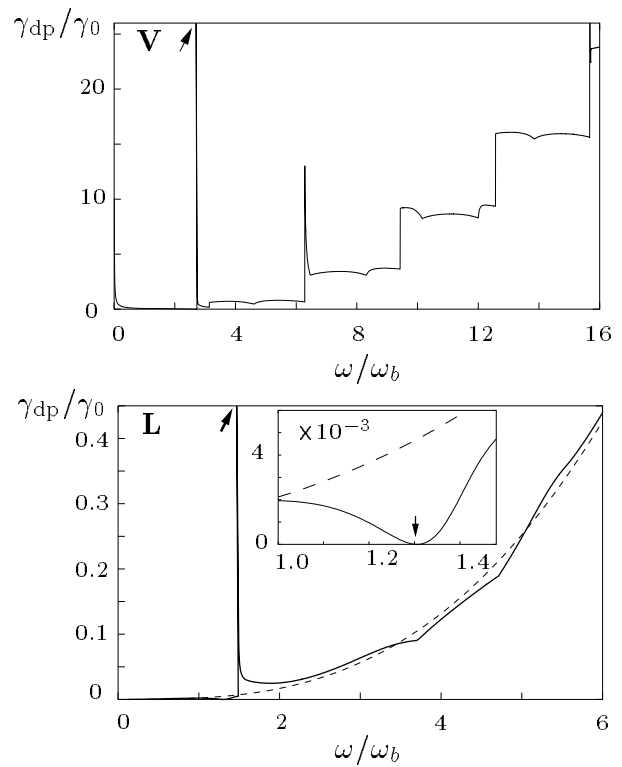,width=0.47\textwidth}
\caption{Inelastic phonon emission rate $\gamma_{\rm dp}(\omega)$ of 
  vertical (V) and lateral (L) double dots in a phonon cavity of width
  $2b$ due to deformation potential.  Phonon-subband quantization
  effects appear on an energy scale $\hbar \omega_b = \hbar c_l/b$
  with the longitudinal speed of sound $c_l$; $\gamma_0$ nominal
  scattering rate (see text). Coupling to {\em flexural} (top) and
  {\em dilatational} modes (bottom, dashed: bulk rate).  Inset:
  Suppression of $\gamma_{\rm dp}(\omega)$ from slab phonons at
  $\omega = \omega_0$ (arrow).}
\label{rates}
\end{center}
\end{figure}

Figure \ref{rates} (top) shows $\gamma_{\rm dp}(\omega)$ in units of the
nominal scattering rate $\gamma_0\equiv T_c^2 \Xi^2/\hbar\rho c_l^4 b$ for
$b=5d$. The phonon subband quantization appears as a staircase in $\gamma_{\rm
  dp}(\omega)$, with the steps corresponding to the onsets of new phonon
subbands. Most strikingly, a van Hove singularity (arrow) occurs due to {\em
  zero phonon group velocity} at that frequency.  This corresponds to a
minimum in the dispersion relation $\omega_{n,q_{\|}}$ for finite ${q}_{\|}$
with preceding {\em negative} phonon group velocity due to the complicated
non-linear structure of the Rayleigh-Lamb equations for the planar cavity.
Additional van Hove singularities occur at higher frequencies (not shown here)
as an irregular sequence that can be considered as `fingerprints' of the
phonon-confinement in a mechanical nanostructure.

In the lateral configuration (Fig.~\ref{polar}), DP couples only to
dilatational modes, in contrast to the vertical case. This is a
trivial consequence of the symmetry of the DP of flexural modes
(Fig.~\ref{fields}\,center) and the fact that in the lateral
configuration the dots are aligned mid-plane. The inelastic rate
including the lowest 4 modes is shown in Fig.~\ref{rates}\,(bottom) in
comparision to the bulk rate. The phonon-subband quantization appears
as cusps in $\gamma_{\rm dp}(\omega)$ for $\omega \gtrsim 2 \omega_b$.
Again, we observe van Hove singularities as fingerprints of the phonon
confinement.

Most strikingly, we find a suppression of the inelastic rate at small
energies $\hbar \omega$, and even its {\em complete   vanishing} at
the energy $\hbar\omega_0\approx 1.3 \hbar \omega_b$ for the lateral
configuration.  As can be seen from (\ref{eq:gamman}) and
(\ref{lambdadp}), the rate $\gamma_0$ vanishes for the frequency
$\omega_0$ defined by the condition $q_t = q_{\|}$.  This is due to a
vanishing divergence of the displacement field $\bs{u}$
(Fig.~\ref{fields}\,right) that implies vanishing of DP $\propto
\nabla \cdot \bs{u}$. Near $\omega_0$, the remaining contribution of
the $n=0$-subband mode is drastically suppressed compared with 3d
phonons (Fig.~\ref{rates}\,bottom, inset).

We checked that this decoupling of electrons and phonons is a generic
feature due to the slab geometry, as are the steps and van Hove
singularities in $\gamma$.  For the piezo--electric interaction,  our
results also reveal a complete vanishing of the inelastic rate
$\gamma_{\rm   pz}$ from dilatational phonons at the energy
$\hbar\omega_0' \approx 0.7 \hbar \omega_b$ where the induced {\em
  polarization field} is zero \cite{DEB02}.  Due to the  symmetry of
the latter, in the vertical and lateral configurations only
dilatational and flexural phonons, respectively, couple to the electrons via
PZ interaction.  Thus, the angular dependence is reversed as compared
to the DP case.

An important consequence of these results is that one can `switch off' either
PZ scattering in the vertical configuration, or DP scattering in the
lateral configuration at a certain energy. Then, the only remaining
electron-phonon scattering is mediated by the other interaction mechanism that
couples the electrons to the flexural modes. For other frequencies $\omega$,
the ratio $\gamma_{\rm pz}/\gamma_{\rm dp} \propto b^2$ can be varied by
changing the slab width. Thus, for very small $b$ the DP interaction dominates
and the proper choice to `switch off' the scattering would be the lateral
configuration, with a small contribution remaining if the material is
piezo--electric, and vice versa. For a GaAs slab of width $2b=1\mu$m and a
tunnel coupling $T_c =10\mu$eV in the lateral configuration tuned to $\hbar
\omega \approx 0.7 \hbar \omega_b$ (no PZ coupling), we obtain a residual
scattering rate of $\gamma_{\rm dp} = 8\cdot 10^{4}$ s$^{-1}$ from DP-coupling
to flexural modes.

The characteristic energy scale for phonon quantum size effects is
$\hbar \omega_b\equiv\hbar c_l/b$.  Using the same parameters as
above, we have $\hbar \omega_b=7.5\mu$eV in GaAs which is within the
limit of energy resolution of recent transport experiments in double
dots \cite{FujetalTaretal}.  A finite slab of lateral dimension $L$
will lead to a broadening of the structures predicted above since the
phonons will aquire a life time $\propto L/c_{l,t}$ which leads to a
smearing of fine structures in $\gamma(\omega)$ on an energy scale
$\hbar c_{l,t}/L\sim 1\mu$eV for $L=10 b$. Finite temperatures yield a
similar broadening on a scale $k_BT$. Therefore, low temperatures (20
mK $\approx 2\mu$eV) are required to resolve the step-like features
and the van Hove singularities in the inelastic current $I_{\rm in}$.
We mention that vertically polarized shear waves (which are also
eigenmodes of the slab) do not couple to electrons via the DP because
the induced local change in volume ($\propto \nabla\cdot\bs{u}$) is
zero.  Fortunately, shear waves also do not change the low frequency
decoupling discussed above because the only mode that is accessable at
low enough energies is the massless mode (linear dispersion) that does
not lead to any piezo-electric polarization field. However, at higher
frequencies shear subbands can contribute to the electron scattering.

The existence of an energy $\hbar\omega_0$ where the electron-phonon
interaction vanishes could be used to suppress decoherence in double
dot qubit systems with the energy difference $\Delta=
\sqrt{\varepsilon^2 +4T_c^2}$ tuned to $\Delta=\hbar\omega_0$.  For
example, using gate voltages to tune $T_c(t) = \Delta/2\sin (\Omega
t)$, $\varepsilon(t) = \Delta \cos (\Omega t)$ as a function of time
defines a one-qubit rotation (`electron from left to right') free of
phonon interaction $\propto \gamma(\Delta)=0$. The condition
$\Delta=\hbar\omega_0$ therefore defines a one dimensional
`dissipation-free manifold' (curve) in the $T_c$--$\varepsilon$
parameter space. In particular, a suppression of decoherence could
then be exploited in adiabatic electron transfers \cite{RB01} or
adiabatic swapping operations \cite{SLM01} in coupled quantum dots.

We recall, however, that corrections to $\gamma$ of 4th and higher
order in the coupling constant (virtual processes) can lead to a small
but final phonon-induced dephasing rate even at
$\Delta=\hbar\omega_0$. Moreover, the dephasing due to spontaneous
emission of photons, although negligeable with respect to the phonon
contribution in second order \cite{FujetalTaretal}, is not altered
unless the whole system is embedded into a photon cavity. Similarily,
plasmons and electron-hole pair excitations in the leads can lead to
dephasing. We suppose that the latter can affect the inter-dot
dynamics of the coupled dots only indirectly via coupling to the leads
and only weakly contribute to dephasing, as is the case for
interactions between dot and lead electrons beyond the Coulomb
blockade charging effect.

Alternatively to suppressing dissipation from phonons at certain
energies, the van Hove singularities of the same system could be used
to enormously enhance the spontaneous emission rate of phonons.  The
electron current $I_{\rm in}\approx -e \gamma(\varepsilon)$ at those
energies is due to strong inelastic transitions.  The term $\left|
  \alpha \pm e^{i \bf{q}_{\|}\bf{d}} \right|^2$ in (\ref{eq:gamman})
determines the angular phonon emission characteristic of the double
dot. Therefore, as a function of energy and orientation, the double
dot can be used as an energy selective phonon emitter with well
defined emission characteristics.

In conclusion, we have found that phonon confinement is a promising
tool for gaining control of dephasing in double quantum dots via
phonons. Once this was achieved, it would be possible to study
systematically dephasing due to other mechanisms such as coupling to
electronic excitations in the leads. In contrast to cavity QED, where
a single, confined photon mode can be tuned on or off resonance with
an atomic transition frequency, the vanishing of spontaneous emission
in phonon cavities is due to zeros in the {\em phonon deformation
  potential or polarization fields} rather than gaps in the density of
states. In addition, we found that phonon emission into characteristic
modes can be enormously enhanced due to van Hove singularities that
could act as strong fingerprints of the phonon confinement if
experimentally detected.

This work was supported by the EU via TMR and RTN projects
FMRX-CT98-0180 and HPRN-CT2000-0144, DFG projects Kr 627/9-1, Br
1528/4-1, and project EPSRC R44690/01. Discussions with R. H. Blick,
T.  Fujisawa, W. G. van der Wiel, and L. P. Kouwenhoven are
acknowledged.


\begin{thebibliography}{10}
    
\bibitem{FujetalTaretal} {T.~Fujisawa {\em et al.}, Science {\bf 282},
    932 (1998); S.~Tarucha {\em et al.}, Microelectr. Engineer. {\bf
      47}, 101 (1999).}
  
\bibitem{Legetal87} {A.~J.~Leggett {\em et al.}, Rev.  Mod. Phys.  {\bf 59}, 1
    (1987).}
  
\bibitem{BK99} T.~Brandes and B.~Kramer, Phys. Rev. Lett. {\bf 83},
  3021 (1999).
  
\bibitem{CR98} {A.~N.~Cleland and M.~L.~Roukes}, Nature {\bf 392}, 160
  (1998).
  
\bibitem{BRWB98} {R.~H.~Blick  {\em et al.}}, Physica B {\bf 249}, 784 (1998).
  
\bibitem{SW92} {J.~Seyler and M.~N.~Wybourne}, Phys. Rev. Lett. {\bf
    69}, 1427 (1992).
  
\bibitem{GRKV97} {A.~Greiner, L.~Reggiani, T.~Kuhn, and L.~Varani},
  Phys.  Rev. Lett. {\bf 78}, 1114 (1997).
  
\bibitem{RK98} {L.~G.~C.~Rego and G.~Kirczenow}, Phys. Rev. Lett. {\bf
    81}, 232 (1998).
  
\bibitem{KW97} {T.~D.~Krauss and F.~W.~Wise}, Phys. Rev. Lett. {\bf
    79}, 5102 (1998).
  
\bibitem{DE00} {M.~S.~Dresselhaus and P.~C.~Eklund}, Adv. Phys. {\bf
    49}, 705 (2000).
  
\bibitem{AK00} {R.~Aguado and L.~Kouwenhoven}, Phys. Rev. Lett. {\bf
    84}, 1986 (2000).
   
\bibitem{Blietal00} {R.~H.~Blick {\em et al.}, Phys. Rev. B {\bf 62},
    17103 (2000).}
 
\bibitem{HN97} {X.~Hu and F.~Nori}, Phys. Rev. Lett. {\bf 79}, 4605
  (1997).
  
\bibitem {MM64Aul73} {T.~Meeker, and A.~Meitzler, in {\em Physical Acoustics}
    (Academic, New York 1964), Vol. 1, Part A; B.~Auld, {\em Acoustic
      Fields and Waves} (Wiley, New York 1973), Vol. 2.}
  
\bibitem{cavityQED} {G. S. Argawal, {\em Fundamentals of Cavity Quantum
      Electrodynamics} (World Scientific, Singapore 1994).}


\bibitem{SN96} {T.~H.~Stoof and Yu.~V.~Nazarov}, Phys. Rev. B {\bf
    53}, 1050 (1996).
  
\bibitem{LL65}{L.~D.~Landau and E.~M.~Lifschitz, {\em Course of
      Theoretical Physics} (Butterworth-Heinemann, Oxford 1986), Vol. 7.}
  
\bibitem{BAMS9594} {N.~Bannov {\em et al.}, Phys. Rev. B {\bf 51}, 9930
    (1995); N.~Bannov {\em et al.}, phys. stat. sol.  (b) {\bf 183}, 131
    (1994).}

\bibitem{DEB02}{S.~Debald, T.~Brandes, and B.~Kramer, unpublished.}
  
\bibitem{RB01}{F.~Renzoni and T.~Brandes, Phys. Rev.  B {\bf 64},
    245301 (2001).}
  
\bibitem{SLM01}{J.~Schliemann, D.~Loss, and A.~H.~MacDonald, Phys.
    Rev. B {\bf 63}, 085311 (2001).}

\end{thebibliography}
\end{document}